\documentclass[doublecol]{epl2} 

\usepackage{graphicx}
\usepackage{amsmath}
\usepackage{amsfonts}
\newcommand{\me}{\mathrm{e}}
\newcommand{\md}[1]{\mathrm{d}#1}

\title{Finite time thermodynamics for a single level quantum dot}

\author{Massimiliano Esposito\inst{1,2} \and Ryoichi Kawai\inst{3} \and Katja Lindenberg\inst{2} \and Christian Van den Broeck\inst{4}}
\shortauthor{M. Esposito \etal}

\institute{               
 \inst{1} Center for Nonlinear Phenomena and Complex Systems, Universit\'e Libre de Bruxelles, Code Postal 231, Campus Plaine, B-1050 Brussels, Belgium.\\
 \inst{2} Department of Chemistry and Biochemistry and BioCircuits Institute, University of California, San Diego, La Jolla, CA 92093-0340, USA.\\
 \inst{3} Department of Physics, University of Alabama at Birmingham, 1300 University Blvd. Birmingham, AL 35294, USA.\\
 \inst{4} Hasselt University, B-3590 Diepenbeek, Belgium.
}


\abstract{
We investigate the finite time thermodynamics of a single-level fermion system interacting with a thermal reservoir through a tunneling junction. The optimal protocol to extract the maximum work from the system when moving the single energy level between an initial higher value and a final lower value in a finite time is calculated from a quantum master equation. The calculation also yields the optimal protocol to raise the energy level with the expenditure of the least amount of work on the system. The optimal protocol displays discontinuous jumps at the initial and final times. 
}

\begin{document}
\maketitle
\section{Introduction} \label{introduction}

The search for the least work-intensive protocols for the extraction or insertion 
of energy into or out of a thermal system has been a major research topic since 
the inception of the laws of thermodynamics. While the regime of quasi-static
transformations, which is described by close-to-equilibrium thermodynamics, is
well understood, many questions remain unsolved when dealing with problems far
from equilibrium. A first question of particular interest deals with
thermodynamic processes taking place in a finite time. This issue has been
the object of detailed investigations in the context of finite time
thermodynamics (FTT)\cite{ftt}. Other developments are related to recent 
progress in nanotechnology and cellular biology, where small systems far
from equilibrium are subject to large thermal fluctuations. Deviations
from average behavior, even rare events, play a significant role in their
behavior. During the past decade, major progress has been achieved toward
understanding and describing the role of fluctuations in such small
non-equilibrium systems. The fluctuation theorem~\cite{evans93,gallavotti95},
the Jarzynski equality~\cite{jarzynski97}, Crook's theorem~\cite{crooks99}, and
the formulation of stochastic thermodynamics~\cite{reviewSeifert} provide a
novel framework to tackle the role of fluctuations in entropy production and
dissipative work far from equilibrium. In addition, exact expressions 
for the irreversible entropy production have also been
derived~\cite{kawai07,jarzynski06,parrondo09,esposito09}.
A third frontline of  research deals with quantum mechanical behavior in FTT. 
As the size of a system is reduced to the nanometer scale, quantum mechanical
properties such as discreteness, quantum coherence, quantum statistics, and 
quantum correlations (entanglement) must be taken into account. We cite in 
particular the thermodynamics of quantum information processing~\cite{alicki04,segawa09}, 
the related quantum heat engines~\cite{alicki04,lloyd97,scully87}, and 
quantum entanglement as a source of canonical typicality~\cite{popescu06}.


One of central questions addressed in FTT is to identify the optimal procedure
to extract the greatest amount of work from a device operating under given
constraints, or in reverse, to cause a device to operate under such constraints
with the minimum injection of work. According to the convention in which
$\mathcal{W}$ is the work
done \textit{on} the system, maximum work extracted or minimum work
injected both correspond to the
minimum $\mathcal{W}$. The question is thus that
of identifying the protocol that involves the minimum amount of work done on
the system. For example, Schmiedl and Seifert~\cite{schmiedl07} considered  the
optimal protocol to relocate a Brownian particle using a laser tweezer.   They
found that the optimal variation of the laser intensity which minimizes the
work done on the system exhibits sudden jumps.  Such singularities in the
optimal protocol may seem surprising, but in fact they turn out to be
generic~\cite{band82,schmiedl08,then08,gomez-marin08}.

In the present letter we address a similar question for a simple quantum
process. We consider a single level quantum system interacting with a heat
bath. By raising or lowering the energy of this level, we can inject work
($\mathcal{W}>0$) or extract work ($\mathcal{W}<0$). The time dependence of
the protocol $\epsilon(t)$ can be controlled externally. Our aim is to find an
optimal protocol, one that minimizes the work done on the system, under the
constraints of given initial and final values $\epsilon_0$ and $\epsilon_1$, and
a fixed total operation time $\tau$. We will specifically consider a quantum
dot and a tunneling junction to a metal lead, the latter playing the role of a
fermionic thermal reservoir.
The detailed analysis of time-dependent phenomena in open quantum system is
extremely complicated. In order to obtain exact analytical and numerical
results, we restrict ourselves to a simple model based on a quantum master
equation. We thus  neglect quantum coherency and entanglement between the
system and the reservoir, but take into account the discreteness of the level
and the proper Fermi-Dirac statistics.


\section{The model} \label{model}

\begin{figure}
\centerline{\includegraphics[width=6.8cm]{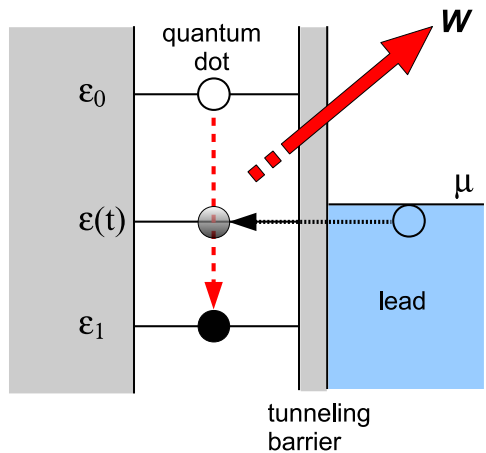}}
\caption{(Color online) The model: a single-level quantum system interacts with
a metallic
thermal reservoir through a tunneling junction.  The level is initially at
$\varepsilon_0$ and in thermal equilibrium.  Work is extracted when the level
is lowered to $\varepsilon_1$.}
\label{fig:model}
\end{figure}

We consider a small quantum system consisting of a single level interacting
with a metallic thermal reservoir through a tunneling junction, as illustrated
in Fig.~\ref{fig:model}.   A quantum dot with a tunneling junction to a lead
provides a realization of such a system, provided the dot has a single energy
level near the Fermi level of the lead, and no direct transitions between levels
in the dot take place upon perturbation.   We further assume that the electrons
thermalize instantaneously upon tunneling to the reservoir, and that the latter
remains in  thermal equilibrium at a constant temperature $T$ at all times.

Upon varying the energy level or the chemical potential, a certain
amount of (positive or negative) energy flows into the system as work done by an
external agent. When the level is lowered (downward process), work is extracted
($\mathcal{W}<0$). On the other hand, when the level is raised (upward process),
work is injected into the system ($\mathcal{W}>0$). The amount of work depends
on the way the energy level is varied. Our goal is to find an optimal way of
varying the energy level, the so-called optimal protocol, such that the maximum
amount of work $-\mathcal{W}$ is extracted from the system or the minimum amount
of work $\mathcal{W}$ is injected.

We describe the time evolution of the quantum state using a master equation for
the occupation probability $p(t)$,
\begin{equation}
 \dot{p}(t)= -\omega_1(t) p(t) + \omega_2(t) [1-p(t)],
\label{eq:master}
\end{equation}
where the $\omega_i$ are transition rates~\cite{harbola06}.  In the wide-band
approximation, these rates are given by:
\begin{subequations}
\label{eq:rates}
\begin{eqnarray}
 \omega_1 &=& \frac{C}{\me^{-\beta[\varepsilon(t)-\mu(t)]}+1} \label{eq:rate1}\\
 \omega_2 &=& \frac{C}{\me^{+\beta[\varepsilon(t)-\mu(t)]}+1} \label{eq:rate2},
\end{eqnarray}
\end{subequations}
where $C$ is a constant.  Noting that raising the energy level is equivalent to
lowering the chemical potential, we introduce an effective energy
$\epsilon(t) \equiv \varepsilon(t)-\mu(t)$.   By measuring time in units
of $C^{-1}$, the master equation~(\ref{eq:master}) thus reduces to the simple
form
\begin{equation}\label{eq:master2}
 \dot{p}(t) = -p(t) + \frac{1}{\me^{\beta \epsilon(t)}+1}.
\end{equation}
We assume that the system is initially in thermal equilibrium,
\begin{equation}
 p(0) = \frac{1}{\me^{\beta\epsilon_0}+1}\, .
\label{eq:p0}
\end{equation}
%

\section{Thermodynamic quantities} \label{thermo}

We next turn to a thermodynamic analysis of the model. We use the convention
that heat entering the system is (like work) positive. The internal energy of
the system at a time $t$ is
\begin{equation}
E(t) = U(t) - \mu N(t) = \epsilon(t) p(t),
\label{internalenergy}
\end{equation}
where
\begin{equation}
U(t) = \varepsilon(t) p(t), \quad N(t) = p(t).
\label{pieces}
\end{equation}
The rate of change in the internal energy,
$\dot{E}$, is the sum of two parts, namely a work flux
 $\dot{\mathcal{W}}$ and a heat flux $\dot{\mathcal{Q}}$,
\begin{subequations}
\begin{eqnarray}
\label{workflux}
\dot{\mathcal{W}} &\equiv& \dot{\epsilon}p = \dot{\varepsilon} p -\dot{\mu} p\\
\label{heatflux}
\dot{\mathcal{Q}} &\equiv& \epsilon\dot{p} = \varepsilon \dot{p} - \mu \dot{p}.
\end{eqnarray}
\end{subequations}
Note that the particle exchange contributes to
the heat flux [last term in Eq.~(\ref{heatflux})]. When the energy level is
below the Fermi
level, the direction of heat flow is opposite to the direction of tunneling.

The net total work and net total heat during the process of duration $\tau$ are
obtained as functionals of the occupation probability,
\begin{subequations}
\begin{eqnarray}
\mathcal{W}[p(\cdot)] &=& \int_{0}^{\tau}  \dot{\epsilon}(t) p(t) \md t \label{eq:W} \\
\mathcal{Q}[p(\cdot)] &=& \int_{0}^{\tau}  \epsilon(t) \dot{p}(t) \md t
\label{eq:Q}.
\end{eqnarray}
\end{subequations}
The resulting total net change in the internal energy is given by the First Law
of thermodynamics,
\begin{equation}
 \Delta E = p(\tau) \epsilon_1 - p(0) \epsilon_0 =\mathcal{W}[p(\cdot)]+\mathcal{Q}[p(\cdot)]\,.
\label{eq:E}
\end{equation}
While work and heat depend on the path of $p(t)$, $\Delta E$ depends only on the
final probability $p(\tau)$ and the given  constraints $\epsilon_0$, $
\epsilon_1$ and $ p(0)$.

\section{Minimizing work} \label{minwork}

\subsection{General Approach} \label{minworka}

Our aim is to find an optimal protocol $\epsilon(t)$ which minimizes the
work $\mathcal{W}$. However, performing a variational analysis directly with
respect to $\epsilon(t)$ is complicated due to the expected discontinuities. 
Instead, we optimize the work with respect to $p(t)$, and identify the
corresponding optimal $\epsilon(t)$ from it.

From the First Law of the thermodynamics, Eq.~(\ref{eq:E}), we find work as a
functional of $p(t)$,
\begin{equation}
\mathcal{W}[p(\cdot)] = \Delta E - \mathcal{Q}[p(\cdot)]\, .
\label{eq:1stlaw}
\end{equation}
Since by definition $p(t)$ is always differentiable, this
expression is well defined.

In order to minimize work, we need to minimize $\Delta E$ and maximize
$\mathcal{Q}$ simultaneously.  However, from Eq.~(\ref{eq:E})  we see
that $\Delta E$ only depends on the final probability $p(\tau)$. Hence, we first
identify the protocol leading to maximum heat $\mathcal{Q}$ for a given value
of $p(\tau)$. In a second step, we perform the optimization with respect to the
final state $p(\tau)$.
To simplify notation, we assume in this section that energy is measured in
units of $kT$.

To find the protocol that maximizes the heat, we express
$\epsilon(t)$ in terms of $p(t)$ and $\dot{p}(t)$ and rewrite Eq.~(\ref{eq:Q})
as
\begin{equation}
 \mathcal{Q}[p(\cdot)]=\int_{0}^{\tau} \mathcal{L}(p,\dot{p}) \md t,
\end{equation}
where
\begin{equation}
 \mathcal{L}\equiv \ln \left [ \frac{1}{p(t)+\dot{p}(t)}-1 \right ] \dot{p}(t).
\end{equation}
The extremum is found via the standard Euler-Lagrange method, leading, after
integration, to
\begin{equation}
 \mathcal{L} - \dot{p} \frac{\partial \mathcal{L}}{\partial \dot{p}} =
\frac{\dot{p}^2}{(p+\dot{p})(1-p-\dot{p})} = K,
\label{eq:K}
\end{equation}
where $K$ is the constant of integration.
Before turning to the solution of this differential equation, we show that it
implies a discontinuity in the protocol $\epsilon(t)$. Eliminating $\dot{p}$ in
Eq.~(\ref{eq:K}) by using the master equation~(\ref{eq:master2}),  the
resulting quadratic equation for $p(t)$ leads to the relation
\begin{equation}
 p(t) =
  \frac{1}{e^{\epsilon(t)}+1}\left [1 \pm \sqrt{K e^{\epsilon(t)}} \right ].
\label{eq:p}
\end{equation}
If one  determines the value of the integration constant $K$ from the initial
condition $p(0)$ assuming $\lim_{t \to 0} \epsilon(t)=\epsilon(0)$, this
relation implies that $K=0$, i.e., that $p(t)$ is the equilibrium distribution
associated with the instantaneous value of the energy. However, one expects that
$p(t)$ will deviate from thermal equilibrium except for an infinitely slow
quasi-static process, so that in general $K\neq 0$.
This apparent inconsistency indicates that $\lim_{t\to 0}\epsilon(t) 
\neq \epsilon(0)$.  In other words, there must be a sudden jump from
$\epsilon_0$ to $\epsilon(0^+)$. By comparing Eq.~(\ref{eq:p0}) to
Eq.~(\ref{eq:p}) at $t=0$, we find the magnitude of the jump,
\begin{eqnarray}
\epsilon(0^+)-\epsilon_0&=&\ln \left [ 1 \pm 2 K\cosh^2\frac{\epsilon_0}{2} \right . \nonumber \\
&\times& \left . \left (1+
\sqrt{1+\frac{1}{K\cosh^2\frac{\displaystyle\epsilon_0}{\displaystyle 2}}}
\right ) \right ].
\end{eqnarray}

Equation~(\ref{eq:p}) also indicates that when $K>0$ there are two
possibilities. The plus sign in $\pm$ leads to an occupation probability $p(t)$
that is larger than that of thermal equilibrium, and corresponds to the
scenario of moving to a higher energy $\epsilon_1 \ge \epsilon_0$. We refer to
these as upward processes.  For downward processes, the lower sign should be
used. Henceforth it should thus be understood that the upper (lower) sign has to
be considered when processes are upward (downward), respectively.

Proceeding with the discussion of Eq.~(\ref{eq:K}), we solve the quadratic
equation for $\dot{p}$, leading to
\begin{equation}
 \dot{p} = \frac{K(1-2p)\mp\sqrt{K^2+4Kp(1-p)}}{2(1+K)} \label{eq:ode}.
\end{equation}
This equation can be solved by separation of the variables $t$ and $p$, leading
to the following  explicit result for the inverse function $t$ as a function of
$p$,
\begin{equation}\label{eq:general}
 t=F[p(t)]-F[p(0)],
\end{equation}
with
\begin{eqnarray}
 F(p) &=& -\frac{1}{2} \ln[p(1-p)] \mp \frac{1}{\sqrt{K}} \arcsin\left(\frac{1-2p}{\sqrt{K+1}}\right) \nonumber\\
&&\hspace{-0.5cm}\pm \frac{1}{2} \text{arctanh}
\left \{ \frac{K(2p-1)[4p(p-1)-K]}{2(K-1)p(p-1)+K}\right \} .
\label{eq:F}
\end{eqnarray}
While in general we will need to proceed with a numerical inversion for the
resulting transcendental equation, an analytically
tractable approximation will be discussed in the next section.
Having thus obtained  the optimal $p(t)$ for a given $K$, we  insert 
this expression in Eq.~(\ref{eq:Q}) to obtain the corresponding heat,
\begin{eqnarray}
&&\hspace{-0.5cm}\mathcal{Q} = \int_0^\tau \epsilon(t) \dot{p} dt = \int_{p(0)}^{p(\tau)}
\epsilon(p) \md p \nonumber \\
&&\hspace{-0.1cm}= \int_{p(0)}^{p(\tau)} \md p \ln \left [
\frac{K+2p-2p^2\pm\sqrt{K^2+4Kp-4Kp^2}}{2p^2} \right ] \nonumber\\
&&\hspace{-0.1cm}= G[p(\tau)]-G[p(0)],
\label{eq:optQ}
\end{eqnarray}
where
\begin{eqnarray}\label{eq:G}
 G(p) &=& p \ln \left [ \frac{K+2p-2p^2 \pm \sqrt{K^2 + 4Kp - 4Kp^2}}{2p^2} \right ] \nonumber \\ &-& \ln(1-p)
 \pm\sqrt{K}\arcsin \left [ \frac{2p-1}{\sqrt{K+1}} \right ] \nonumber \\
&\mp& \text{arctanh}\left ( \frac{2p-2-K}{\sqrt{K^2 + 4Kp - 4Kp^2}} \right) .
\end{eqnarray}
Finally, we need to optimize the resulting work, given in Eq.~(\ref{eq:1stlaw}),
with respect to $p(\tau)$, as explained earlier.  Since $p(\tau)$ is uniquely
determined by $K$, it suffices to numerically optimize the expression with
respect to  $K$. 

We note some useful symmetries that arise from the fact that the optimal
protocol to lower the energy level (with a resulting extraction of
work from the system) is the mirror image in time of the optimal protocol to
raise the energy level (associated with an injection of work into the system).
Denoting the optimal protocol for upward processes from $\epsilon_0$ to
$\epsilon_1 > \epsilon_0$ by $\epsilon_\uparrow(t)$ and the optimal protocol for
downward processes from $-\epsilon_0$ to $-\epsilon_1$ by
$\epsilon_\downarrow(t)$ and the associated occupation probabilities by
$p_\uparrow(t)$ and $p_\downarrow(t)$ one easily finds that
\begin{eqnarray}
&& \epsilon_\uparrow(t) + \epsilon_\downarrow(t) = 0 \label{eq:sym_e},\\
&& p_\uparrow(t)+p_\downarrow(t)=1. \label{eq:sym_p}
\end{eqnarray}
These symmetries also imply symmetries for the minimum work and the associated
heat,
\begin{eqnarray}
 W_\uparrow -  W_\downarrow &=& \epsilon_1 - \epsilon_0, \label{eq:sym_W}\\
 Q_\uparrow &=& Q_\downarrow. \label{eq:sym_Q}
\end{eqnarray}
These symmetries can be thought of as electron-hole symmetries. They are
particularly useful when the initial and final levels are symmetric with
respect to the Fermi level, i.e., when $\epsilon_1 = - \epsilon_0$.

\subsection{The high temperature regime} \label{minworkb}

The mathematical expressions for the general case derived  in the previous
subsection are rather complicated.   However, the  functions (\ref{eq:F}) and
(\ref{eq:G}) simplify  in the high-temperature limit, allowing us to find  the
optimal protocol and its properties in full analytical detail.

First, we introduce an effective energy level $\eta(t)$ defined by
\begin{equation}
 p(t) = \frac{1}{e^{\eta(t)}+1},
\label{eq:eta}
\end{equation}
with $\eta(t)$ implicitly defined via Eq.~(\ref{eq:p}).
Next, we consider Eqs.~(\ref{eq:F}) and (\ref{eq:G}) as functions of $\eta$.
Since $\eta(t) \ll 1$ and $\epsilon(t) \ll 1$ in the high temperature limit, we
keep only lowest order terms. Noting from Eq. (\ref{eq:K}) that $\sqrt{K}$ is of
the same order as $\epsilon$ and $\eta$, we find that Eqs.~(\ref{eq:eta}),
(\ref{eq:p}), (\ref{eq:F}) and (\ref{eq:G}) simplify as follows:
\begin{eqnarray}
 p(t) &=& \frac{1}{2} - \frac{\eta(t)}{4}\\
 \epsilon(t) &=& \eta(t) \pm 2 \sqrt{K}\\
 F(\eta(t)) &=& \pm \frac{\eta(t)}{2\sqrt{K}}\\
 G(\eta(t)) &=& \mp \frac{1}{2} \sqrt{K} \eta(t) - \frac{1}{8}\eta^2(t)\,.
\end{eqnarray}
Solving Eq.~(\ref{eq:general}), we find $\eta(t)=\epsilon_0 \pm 2 \sqrt{K}t$ and
hence the optimal protocol reads
\begin{equation}\label{eq:highT}
 \epsilon(t) = \epsilon_0 \pm 2 \sqrt{K}(t+1).
\end{equation}
The work is minimum for
\begin{equation}
K = \frac{1}{4} \left ( \frac{\epsilon_1-\epsilon_0}{\tau+2} \right )^2\;.
\label{eq:optK_highT}
\end{equation}
The optimal work and associated heat thus become
\begin{subequations}
\begin{eqnarray}
 W&=&\frac{(\epsilon_1-\epsilon_0)[8-4\epsilon_0+\tau(4-\epsilon_1-\epsilon_0)]}{8(\tau+2)}
\label{eq:optW_highT},\\
 Q&=&\frac{\tau(\epsilon_0^2-\epsilon_1^2)}{8(\tau+2)}\, .
\label{eq:optQ_highT}
\end{eqnarray}
\end{subequations}

From Eq. (\ref{eq:highT}) we find that the initial and final energy jumps are
given by $\pm 2 \sqrt{K}$. The size of the jumps increases as the deviation from
the quasi-static limit (measured by $K$) increases.  In between the jumps, the
optimal protocol raises/lowers the level linearly with time (but we stress that
this linear dependence only applies to the high temperature regime). Note that
the results of the high temperature approximation satisfy the symmetry
relations Eqs.~(\ref{eq:sym_e})-(\ref{eq:sym_Q}). Here for the
case with the symmetry $\epsilon_0= - \epsilon_1$
there is no net heat flow. In this case all the work is converted into
internal energy.

\begin{figure}
\includegraphics[width=3.0in]{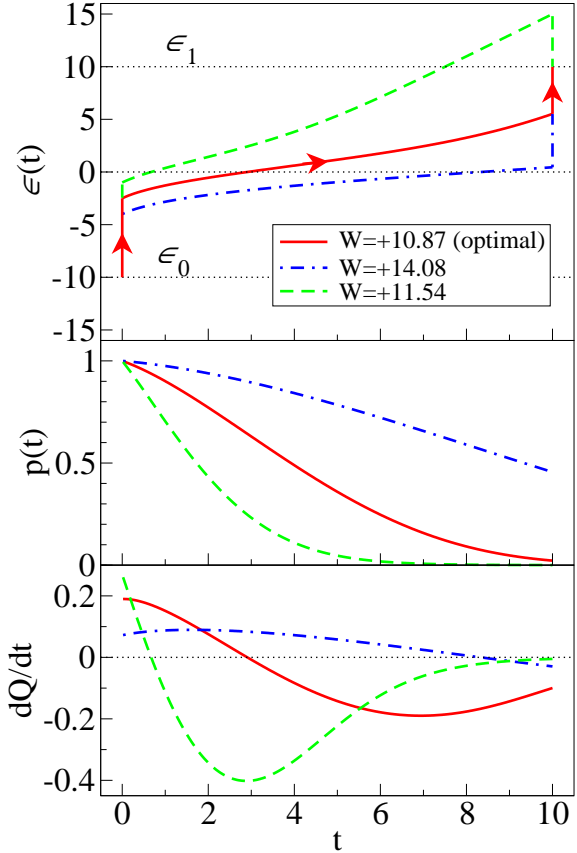}
\caption{(Color online) Protocol (top), occupation probability (middle), and
heat flux (bottom).  The energy level is raised from $\epsilon_0=-10\, kT$ to
$\epsilon_1=+10\, kT$ during time $\tau=10$. The red solid lines indicate
optimal protocol with minimum work $W=+10.87\, kT$.  A protocol with a large
initial jump (green dashed lines) and another protocol with a small initial jump
(blue dot-dahed lines) result in higher work $W=+16.79\, kT$ and $W=+11.16\,
kT$, respectively.}
\label{fig:upslow}
\end{figure}

\begin{figure}
\centerline{\includegraphics[width=3.0in]{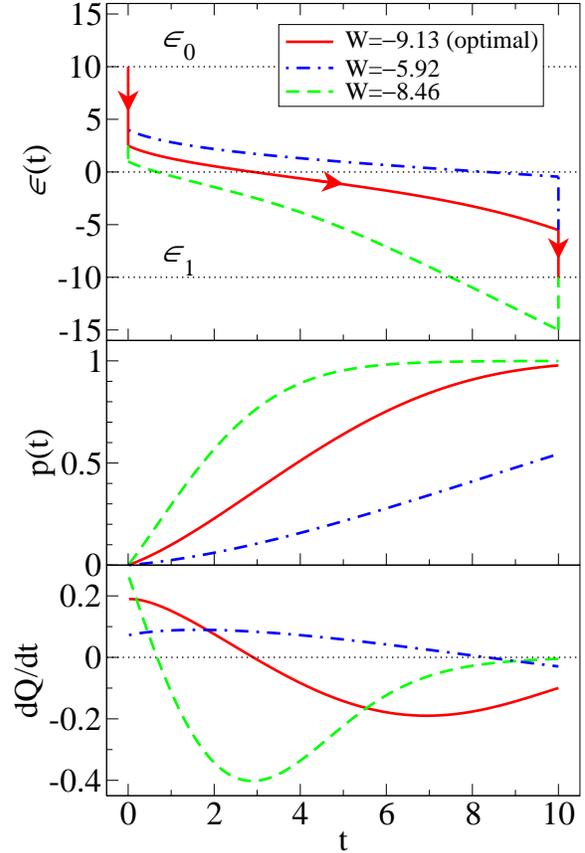}}
\caption{(Color online) Protocol (top), occupation probability (middle), and
heat flux (bottom).  The energy level is lowered from $\beta \epsilon_0=+10\,
kT$ to $\epsilon_1=-10\, kT$ during time $\tau=10$. The red solid lines indicate
optimal protocol with minimum work $ W=-9.13\, kT$.  A protocol with a large
initial jump (green dashed lines) and another protocol with a small initial jump
(blue dot-dahed lines) result in higher work $W=-8.58\, kT$ and $W=-5.92\, kT$,
respectively.  All three values of the work are related to the work in Fig.
\ref{fig:upslow} through the electron-hole symmetry (\ref{eq:sym_W}).}
\label{fig:downslow}
\end{figure}

\section{Results}
\label{results}

In this section we present results  for the optimal protocol obtained via 
numerical solution for a number of representative cases.
First we consider the situation where the energy level is raised from
$\epsilon_0=-10\, kT$ to $\epsilon_1=+10\, kT$ during a total available time
$\tau=10$. This is the situation in which work is done on the quantum dot. 
The chosen parameter values guarantee that the initial energy level is well 
below, and the final level well above, the Fermi level. 
The top panel in Fig.~\ref{fig:upslow} shows the optimal
protocol. We also include two other protocols which do not have the optimal
value of the initial jump, that is, ones corresponding to non-optimal values of
$K$. The middle and bottom panels show the corresponding occupation
probabilities and heat current. When the initial jump is ``too small"
(dot-dashed lines), the level stays mostly below the Fermi level and thus the
system receives heat from the reservoir. While this favors the reduction of
work, there is not enough time for electrons, being below the Fermi most of the
time, to tunnel, thus costing a large amount of work when raising the electrons
during the final jump.
On the other hand, when the initial jump is too large (dashed line), the
electrons quickly escape and thus almost no work is required at the final jump. 
However, the rapid tunneling induces a large heat current to the reservoir,
increasing the work during the process.  The optimal protocol (solid line)
guarantees that the level is most likely empty before the final jump, with only
a small outbound heat current.  Note that the heat initially flows into the
system, compensating for part of the outgoing heat after $\epsilon(t)$ crosses
the Fermi level.

Figure \ref{fig:downslow} shows downward processes where the energy level is
lowered from $\epsilon_0=+10\, kT$ to $\epsilon_1=-10\, kT$ over the same period
of time $\tau=10$. In this scenario, work is extracted from the quantum
dot. Three different cases parallel to the upward cases of
Fig.~\ref{fig:upslow} are illustrated. All the data shown in
Fig.~\ref{fig:downslow} confirm the electron-hole symmetries surrounding
Eqs.~(\ref{eq:sym_e})--(\ref{eq:sym_Q}).

\begin{figure}
 \centerline{\includegraphics[width=3.3in]{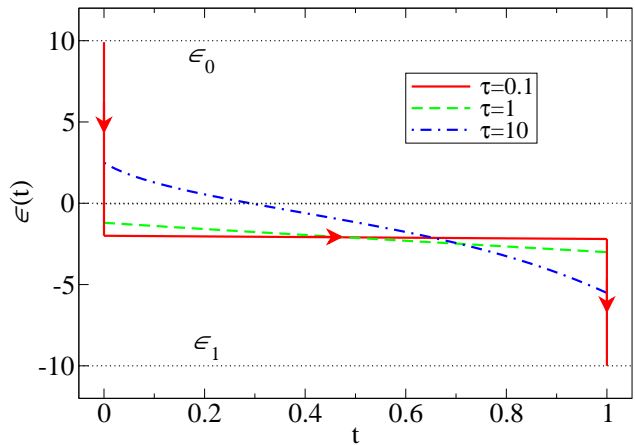}}
 \caption{(Color online) The optimal protocols for three different processing
times, $\tau=0.1$ (solid line), $\tau=1$ (dashed line), and $\tau=10$
(dash-dotted line).
The energy level is lowered from $\beta \epsilon_0=+10\, kT$ to
$\epsilon_1=-10\, kT$ during time $\tau$.}
\label{fig:tau}
\end{figure}

Next, we reduce the time of operation.  Figure~\ref{fig:tau} compares optimal
protocols for slow ($\tau=10$), intermediate ($\tau=1$) and fast ($\tau=0.1$)
processes of work extraction from the quantum dot.  As $\tau$ decreases, the
initial and final jumps become larger. When the electrons have almost no time to
tunnel into the system, the optimal protocol becomes nearly a step function,
i.e., a jump-stay-jump process.

\begin{figure}
 \centerline{\includegraphics[width=3.3in]{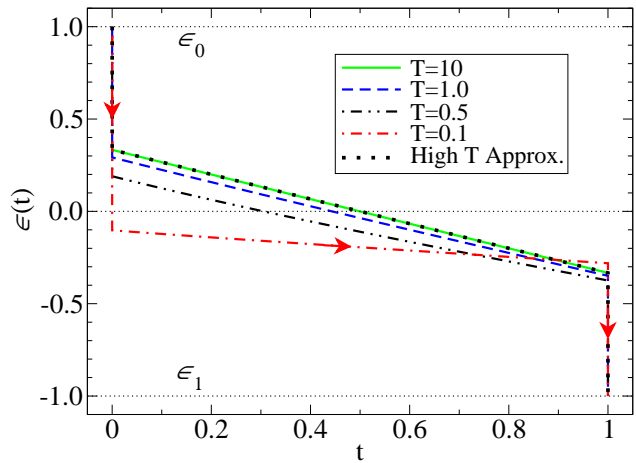}}
 \caption{(Color online) The optimal protocols for four different temperatures,
$T=10$ (green solid line), $T=1$ (blue dashed line), $T=0.5$ (black
dash-double-dotted line), and $T=0.1$ (red dash-dotted line).  The energy level
is lowered from $\epsilon_0=1$ to $\epsilon_1=-1$,}
\label{fig:T}
\end{figure}

Finally, Fig.~\ref{fig:T} shows the optimal protocols for work extraction
at various temperatures. As temperature is increased, the optimal protocol
becomes symmetric with respect to the Fermi level, in good agreement with the
high temperature approximation, Eq.~(\ref{eq:highT}).

\section{Discussion} \label{discussion}

Discontinuities in the protocol that minimizes the work on a device operating
under given constraints may seem to be surprising. However, a simple
phenomenological argument explains the initial and final jumps in the optimal
protocols, starting from an analysis of the low temperature case. The key points
to keep in mind are: (i) moving an empty level requires no work,
(ii) tunneling at the Fermi level carries no heat, and (iii) the direction of
heat flow changes at the Fermi level.   Consider an upward process at
$T=0$. Since tunneling is not possible below the Fermi level, the same amount of
work is required  to raise the level to the Fermi level regardless of the
protocol. Hence, the instantaneous jump to the Fermi level is preferred since it
leaves maximum time for electrons to subsequently tunnel out. After the jump, it
is clear that the optimal protocol must keep the level  infinitesimally above
the Fermi level until the final time. In this way, heat transfer to the
reservoir is avoided. During this period no work is done, and the population in
the energy level is reduced to close to zero without heat transfer. At the end,
the level jumps up to its final value with almost no work.

When the temperature is finite, tunneling is possible even below the Fermi
level. Note that heat flux is inward into the system when the electrons tunnel
out below the Fermi level, which helps reduce the work.  On the other hand,
the tunneling rate is small, which increases the work at a later time.  The
optimal protocol now includes an initial jump to a level slightly below the
Fermi level. At this stage, heat flows into the system. Next,  the energy level
moves slowly above the Fermi level.  As electrons tunnel out, heat flows to the
reservoir, which compensates the initial heat gain.  At the high temperature
limit, the loss and gain of the heat are exactly balanced and no net heat flows
to the reservoir during the process. All results shown in the previous section
are consistent with this phenomenological argument.

We close with a critical  discussion concerning the discontinuities in the
optimal protocol. The jumps have been derived in the context of the master
equation, which is valid only on a coarse grained time scale. 
Furthermore, we know from the time-energy uncertainty principle that an
instantaneous jump in energy level would redistribute  electrons over all energy
levels including continuous states, i.e., the single level model cannot hold in
this strict jump limit. Therefore,
the discontinuities in the optimal protocols identified here should be
interpreted as rapid but continuous changes of the energy level. The
corresponding typical time $\delta t$ must be much shorter than  the tunneling
time ($\delta t \ll 1$). On the other hand, the single level model will remain
valid only when $\delta t \gg 1/\Delta \omega$, where $\Delta \omega$ is the gap
between the energy levels. Even if the chemical potential is changed, we
have assumed that the reservoir remains in thermal equilibrium.  Hence,
$\delta t$ must be longer than the relaxation time of the reservoir.
These conditions can be satisfied if the tunneling rate is small, which is 
also a requirement for the validity of the master equation itself.

\acknowledgments

M. E. is supported by the FNRS Belgium (charg\'e de recherches) and
by the Luxemburgish government (Bourse de formation recherches).
This research is supported in part by the
National Science Foundation under grant PHY-0855471.



\end{document}